\documentclass[showpacs,preprintnumbers,amsmath,amssymb]{revtex4}
\usepackage{graphicx}
\usepackage{dcolumn}
\usepackage{bm}
\begin{document}
\title{Quantization of gauged Floreanini-Jackiw chiral boson with Faddeevian anomaly}


\author{Shahin Absar, Sanjib Ghoshal, Anisur Rahaman}
\email{1. anisur.rahman@saha.ac.in, 2. manisurn@gmail.com
(Corresponding Author)} \affiliation{Hooghly Mohsin College,
Chinsurah, Hooghly - 712101, West Bengal, India}
\date{\today}
\begin{abstract}
\begin{center}Abstract
\end{center}
We consider the gauged model of Siegel type chiral boson with a
Lorentz non-covariant mass-like term for the gauge fields which is
found to be equivalent to the chiral Schwinger model with
Faddeevian anomaly when it is described in terms of
Floreanini-Jackiw type chiral boson. We carry out the quantization
of gauge non-invariant version this model in both the Lagrangian
and Hamiltonian formulation. The quantization of the
gauge-invariant version of this model in the extended phase space
also has been carried out in the Lagrangian formulation. The
gauge-invariant version of this model in the extended phase space
is found to map onto the physical phase space with the appropriate
gauge fixing condition. BRST symmetry associated with this model
has been studied with different gauge fixing terms. It has been
shown that the same model shows off-shell as well as on-shell BRST
invariance depending on the choice of gauge fixing term.

\end{abstract}
 \maketitle
\section{Introduction}
Chiral boson is relevant for the understanding of several field
theoretical models and quantization of chiral boson is interesting
in the regime of lower-dimensional field theory. It appeared
initially in the study of heterotic string theory \cite{ST1, ST2,
ST3, ST4}. The study of quantum Hall effect too got important
input from chiral boson \cite{HAL1, HAL2}. We have found an
interesting description of chiral boson in the pioneering work of
Seigel \cite{SIG}. An alternative illustration of chiral boson was
offered by Srivastava in the article \cite{PPS}. The Lagrangian of
chiral boson was formulated with the second-order time derivative
of the field in both these illustrations \cite{SIG, PPS}, however,
there is a foundational difference in the process of
implementation of chiral constraint through the Lagrange
multiplier. In the original version of Siegel, the chiral
constrain was inserted in a quadratic form, on the contrary, in
the description of Srivastava it was placed in a linear form. An
innovative and sophisticated interpretation of chiral boson also
came from the description of Floreanini and Jackiw \cite{FJ}.

An illuminating illustration towards quantization of free chiral
boson was carried out in the article \cite{JSON}. Extension of
 free chiral boson for the different purpose have been made in
the articles \cite{SIG, PPS, FJ, JSON, AMOR, AMOR1, BPM, RPM}. The
study of free chiral boson is still acquiring a prominent position
in the active field of research. A very recent development towards
the BFV quantization of the free chiral boson along with the study
of Hodge decomposition theorem in the context of conserved charges
has been perused in the article \cite{BPM} In the article
\cite{RPM}, and an application of augmented super field approach
to derive the off-shell nilpotent and absolutely anti-commuting
anti-BRST and anti-co-BRST symmetry transformations for the BRST
invariant Lagrangian density of a free chiral boson has been
executed in a significant manner. An equivalence between gauge
invariant and gauge non-invariant solution of gauged chiral boson
with Faddeevian anomaly with BRST quantization is made by us in
\cite{ARSG}.

Study of interacting chiral boson has also received a great deal
of attention. A spontaneous generalization of a free chiral boson
is to  take into consideration its interaction with the gauge
field, and this interacting  field theoretical model is commonly
known as gauged model of chiral boson. In the article \cite{BEL}
the interacting theory  of chiral boson was initiated and
described in detail by Bellucci, Golterman, and Petcher. The basic
foundation of generalization towards taking into account the
interaction with the gauge field however is laid in the Siegel's
construction of free chiral boson. So the theory of interacting
chiral boson for the Faddeev-Jackiw (FJ) type description appears
like a natural extension when the description of free FJ type
chiral boson became accessible from the article \cite{FJ}. An
ingenious illumination was brought forward by Harada \cite{KH} in
that context. The illuminating extension of Harada \cite{KH}
concerning interacting chiral boson based on FJ type kinetic term
receives a great deal of attention \cite{BAR, BAL, ARC, ABREU,
EUNE, SG}, although this theory of interacting chiral boson was
not derived from the iterating theory of chiral boson that
developed in the article \cite{BEL}. Actually Harada obtained it
from Jackiw-Rajaraman (JR) version chiral Schwinger model
\cite{JR} with the fruitful use of chiral the constraint in the
phase space this theory. So there laid a missing link between
these two types of interacting gauged chiral boson. An attempt
towards the search for a link was therefore a natural pursuit
which we have attempted to link up in our article\cite{ARS}. In
the article \cite{ARS} we have shown that gauged model of chiral
boson with FJ type chiral boson remains in disguise in the gauged
model of chiral boson described in \cite{BEL} with Siegel type
chiral boson. However the gauged model of chiral boson associated
with FJ type description of free chiral boson as extracted out by
us from the description of the article, \cite{BEL} is identical to
the chiral Schwinger model with the standard Jackiw-Rajaraman type
of anomaly when it described in terms of chiral boson \cite{KH}.

If we look towards the background it reveals that chiral
generation of Schwinger model \cite{SCH} due to Hagen \cite{HAG}
had to suffer for a long period due to the non-unitarity problem.
That problem was eradicated taking anomaly into consideration by
Jackiw and Rajaraman \cite{JR}. However, this is not the only
possibility to get out of this unitarity problem which we found in
the article of Mitra \cite{PM}. In the article \cite{PM}, he has
ingeniously shown that the unitarity problem of chiral Schwinger
model can be eradicated with a special type of anomaly termed as
Faddeevian anomaly by Mitra since Gauss law commutator of the
model with itself along with this anomaly remains non-vanishing.
So the formulation of a gauged chiral boson with FJ type free
chiral boson which would be consistent with chiral Schwinger model
with the Faddeevian anomaly would certainly be of interest. A
description of chiral Schwinger model with Faddeevian anomaly in
terms of chiral boson is found in \cite{PMS}. However, in this the
article, it was described in an ad hoc manner: the direct link of
it with its mother version \cite{PM} was not transparent. In
\cite{AR1}, we have made an attempt to shown that this model also
can be generated from the model described by Mitra in \cite{PM}
imposing a chiral constraint in a similar way Harada \cite{KH} did
it for the usual Chiral Schwinger mode. So a natural extension
which would be of interest and instructive as well to investigate
whether the root of chiral Schwinger model with the Fadeevian
anomaly described in terms of a chiral boson is inlaid in the
discrimination of gauged chiral boson in the pioneering work
\cite{BEL} which was done for Harada's version in \cite{ARS}. So,
by all means, the investigation towards the exploration of that
and the systematic study towards determination of spectrum in the
Lagrangian formulation, its correlation with the hamiltonian
formulation and the gauge and BRST invariance of this model as
well is attempted in this article.

It would be beneficial from another point of view too because a
careful look over the previous studies reveals that the gauged
model of chiral boson has a crucial link with anomaly \cite{JR,
KH, PM, PMS, ARC, ARPL, SM1, SM2, ARUP, ARAN1, ARAN2,KH1, KH2},
because it has been found that gauged model of a chiral boson is
crucially connected to chiral Schwinger model and that very chiral
Schwinger model got secured from the non-unitarity problem when
the anomaly was taken into consideration by Jackiw and Rajaraman
\cite{JR}. In this respect, the recent chiral generation of The
Thirring-Wess model is of worth mentioning \cite{ARAN1, ARAN2}.

This article is organized as follows. Sec. II contains the
formulation of gauged Floreanini-Jackiw type chiral boson that
corresponds to the Faddeevian anomaly. Sec. III is devoted to the
determination of theoretical spectra of this model in the
Lagrangian formulation. In Sec. IV, a transparent description of
the evaluation of the theoretical spectra of this model in the
Hamiltonian formulation is presented. Sec. V holds a description
of the determination of the theoretical spectrum of the
gauge-invariant version of the theory in the Lagrangian
formulation. In Sec. VI, an attempt is made to map the gauge
symmetric version of the theory in the extended phase space onto
the gauge non-invariant version of it in the usual Phase space.
Sec. VII is devoted to the study of BRST symmetry of this model.
And the final Sec. VIII contains a brief summary and discussion
over the work.

\section{Formulation of gauged Floreanini-Jackiw type chiral boson
that corresponds to Faddeevian anomaly} A gauged model of Siegel
type chiral boson which resembles chiral Schwinger model with
Jackiw-Rajaraman's one parameter class of regularization was
discussed in \cite{ARS}. A natural extension that trails with is
the gauge model of of Siegel type chiral boson which would be
consistent with   gauged Lagrangian of Floreanini-Jackiw type
chiral boson with Faddeevian anomaly. To formulate that let us
proceed with the following Lagrangian containing a Lorentz
non-covariant mass like term for the gauge field.
\begin{eqnarray}
L&=& \int dx[
\frac{1}{2}(\dot{\phi}^{2}-\phi'^{2})+e(\dot{\phi}+\phi')(A_{0}-A_{1})
+\frac{\Lambda}{2}[(\dot{\phi}
-\phi')+e(A_{0}-A_{1})]^{2}\nonumber \\
&+&\frac{1}{2}(\dot{A}-A'_{0})^{2}+\frac{1}{2}e^{2}(A_{0}+A_{1})(A_{0}-3A_{1})]
\label{SIGCB}
\end{eqnarray}
This Lagrangian is constructed following the article \cite{BEL}.
Here $\phi$ represents scalar field. The gauge field has two
components $A_0$ and $A_1$ in $(1+1)$ dimensional space time.
$\Lambda$ is a lagrange multiplier field. The coupling constant
$e$ has the dimension of mass. The model is holding a Lorentz
non-covariant masslike term for the gauge field. It looks strange
but what it renders is interesting and does not violate physical
Lorentz invariance. It also establishes whether the root of chiral
bosonized version of chiral Schwinger model with Faddeevian
anomaly is in laid in the description of gauged chiral boson
\cite{BEL} or not. We are going to illustrate it in details in
this section. We need to compute the canonical momenta
corresponding to the fields $A_0$, $A_1$, $\phi$ and $\Lambda$.
\begin{equation}
\frac{\partial L}{\partial\dot{A}_0}=\pi_0\approx 0 \label{MOM1}
\end{equation}
\begin{equation}
\frac{\partial
L}{\partial\dot{A}_1}=\pi_1=(\dot{A}_1-A'_0)\label{MOM2}
\end{equation}
\begin{equation}
\frac{\partial
L}{\partial\dot{\phi}}=\pi_{\phi}=(1+\Lambda)\dot{\phi}-\Lambda\phi'
+e(1+\Lambda)(A_{0}-A_{1})\label{MOM3}
\end{equation}
\begin{equation}
\frac{\partial L}{\partial\dot{\Lambda}}=\pi_{\Lambda}\approx
0\label{MOM4}
\end{equation}
A Legendre transformation $H= \pi_0\dot{A}_0 + \pi_1\dot{A}_1 +
\pi_{\phi}\dot{\phi} + \pi_{\Lambda}\dot{\Lambda} -L$  with the
use of the expression of momenta (\ref{MOM1}), (\ref{MOM2}),
(\ref{MOM3}), (\ref{MOM4}) leads to the canonical Hamiltonian
\begin{eqnarray}
H_{C}&=& \int dx {\cal H}_{C}= \int dx[
\frac{1}{2}\pi_{1}^{2}+\pi_{1}A'_{0}+\pi_{\phi}\phi'
-e(\pi_{\phi}+\phi')(A_{0}-A_{1})\nonumber \\
&+& 2e^2 A_{1}^{2}+ \frac{1}{2(1+\Lambda)}(\pi_\phi-\phi')^2]
\end{eqnarray}
The equations (\ref{MOM1}) and (\ref{MOM4}) do not contain any
time derivative so these are the primary constraint of the theory.
The preservation of these constraints lead to new constraints.
Repeating this preservation criteria of the usual constraint and
the forth coming secondary constraints we find that the phase
space of the system is embedded with the following six
constraints.
\begin{equation}
\Omega_{1}=\pi_{0}\approx 0
\end{equation}
\begin{equation}
\Omega_{2}=\pi_{\Lambda}\approx 0
\end{equation}
\begin{equation}
\Omega_{3}=\pi'_{1}+e(\pi_{\phi}+\phi')\approx 0
\end{equation}
\begin{equation}
\Omega_{4}=\pi_{\phi}-\phi'\approx 0
\end{equation}
\begin{equation}
\Omega_{5}=A_0 + A_1 \approx 0
\end{equation}
\begin{equation}
\Omega_{6}=\Lambda-f\approx 0
\end{equation}
Therefore, the generating functional of the theory can be written
down as
\begin{eqnarray}
Z&=&\int |det[\Omega_l, \Omega_m]^{\frac{1}{2}}| dA_{1}d\pi_{1}
d\phi d\pi_{\phi} d\Lambda d\pi_{\Lambda}dA_{0}d\pi_{0}e^{i\int
d^2x(\pi_{1}\dot{A_{1}}+\pi_{\phi}\dot{\phi}+\pi_{\Lambda}\dot{\Lambda}
+\pi_{0}\dot{A_{0}}-{\cal H}_{C})}\nonumber\\
&&\times
\delta(\Omega_{1})\delta(\Omega_{2})\delta(\Omega_{3})\delta(\Omega_{4})
\delta(\Omega_{5})\delta(\Omega_{6})
\end{eqnarray}
The subscripts $l$ and $m$ runs from $1$ to $6$. After
simplification by the use of gaussian integral we land on to
\begin{equation}
Z=\int d\phi dA_{1}dA_{0} e^{i\int d^2x {\cal L}_{CH}}
\end{equation}
where
\begin{equation}
{\cal L}_{CH}
=\dot{\phi}\phi'-\phi'^{2}+2e\phi'(A_{0}-A_{1})-2e^2A^{2}_{1}
+\frac{1}{2}(\dot{A_{1}}-A'_{0})^{2} \label{FJCB}\end{equation}
 So it is now evident that the
Lagrangian (\ref{SIGCB}) is the appropriate gauged Lagrangian of
Siegel type Chiral Boson that corresponds to the gauged chiral
boson with Floreanini-Jackiw type chiral boson which can generated
from chiral Schwinger model with Faddeevian anomaly (\ref{FJCB})
introducing a chiral constraint in the phase space of the theory
\cite{ARC}. We will now turn to find out the theoretical spectrum
of this system in the Lagrangian formulation because what a model
with Lorentz non-covariant structure offers would be of interest.

\section{Determination of theoretical spectrum in the Lagrangian
formulation} The gauged Lagrangian  density of chiral boson with a
the Mitra type faddeevian anomaly  anomaly is given by
\begin{equation}
{\cal L}_{CH}=\dot{\phi}\phi'-\phi'^2+2e\phi'(A_{0}-A_{1})-2e^2
A_1^2 +\frac{1}{2}(\dot{A}_1-A'_0)^2 \label{LCH}
\end{equation}
This Lagrangian is not gauge symmetric in its usual phase space
and  the structure of this Lagrangian is not Lorentz
non-covariant.  We will proceed to obtain the theoretical spectrum
of the system described in a gauge non-invariant way in the
Lagrangian formulation. Using Euler-Lagrangian equations we obtain
the following three equations of motion corresponding to the field
$\phi$, $A_1$ and $A_0$
\begin{equation}
\dot{\phi}'-\phi''+e(A_{0}-A_{1})=0, \label{NEQ1}
\end{equation}
\begin{equation}
-\ddot{A_{1}}+\dot{A_{0}}'-2e\phi'-4e^{2}A_{1}=0,\label{NEQ2}
\end{equation}
\begin{equation}
-A''_0+\dot{A_{1}}'+2e\phi^{'}=0. \label{NEQ3}
\end{equation}
To solve these coupled differential  equations (\ref{NEQ1}),
(\ref{NEQ2}) and (\ref{NEQ3}) we introduce the ansatz for the
fields $A_{\mu}$and $\phi$ as
\begin{equation}
A_{\mu}=\frac{1}{4e^{2}}\tilde{\partial}F, \label{AN1}
\end{equation}
\begin{equation}
\phi=\frac{1}{4e}F.\label{AN2}
\end{equation}
Equation (\ref{NEQ1}) gives
\begin{equation}
\Box F = -\frac{1}{4e^2}(\dot{A}_1 - A'_0)\label{SOLF}
\end{equation} Plugging in the ansatz (\ref{AN1}) and (\ref{AN2}) in the
equations (\ref{NEQ1}), (\ref{NEQ2}) and (\ref{NEQ3}) and using
(\ref{SOLF})the following solution of $F$ is obtained:
\begin{equation}
(\Box+4e^{2})\Box{F}=0, \label{OS}
\end{equation}
with the restriction
\begin{equation}
A_0 + A_1 = 0.
 \end{equation}
Note that the equation (\ref{OS}) is Lorentz co-variant although
the Lagrangian was not manifestly Lorentz covariant. The equation
(\ref{OS})indicates that the physical spectrum contains a massive
boson only wit mass $2e$. At a first glance it seems that this
restriction is put by hand on the dynamics of fields $A_0$ and
$A_1$, but that is not the case and it will be transparent if we
look care fully the Hamiltonian analysis of the model which has
already been studied in \cite{PM}. In the following section
however we will describe the Hamiltonian formulation in a coherent
manner to ensure that the restriction is already there in  the
phase space of this constrained system and the spectrum obtained
in the Lagrangian formulation agrees with the Spectrum in the
Hamiltonian formulation. It will  make this article self contained
too.
\section{A transparent Description of theoretical spectrum in the Hamiltonian
formulation} To start with we should mention that main result as
we are going to produce is already known from the articles
\cite{PM, PMS}. To make this article self contained and to make an
easy comparison of the Lagrangian and Hamilonian formulation in
connection with the theoretical spectrum we are furnishing  it in
a desired cohered  manner. To obtained the theoretical spectrum in
the Hamiltonian formulation we calculate the momenta corresponding
to the fields $A_0$, $A_1$ and $\phi$ from the Lagrangian density
(\ref{LCH}) using the standard definition of momenta.
\begin{equation}
\frac{\partial L_{CH}}{\partial\dot{A}_0}=\pi_0\approx 0,
\label{TMOM1}
\end{equation}
\begin{equation}
\frac{\partial
L_{CH}}{\partial\dot{A}_1}=\pi_1=(\dot{A}_1-A'_0),\label{TMOM2}
\end{equation}
\begin{equation}
\frac{\partial
L_{CH}}{\partial\dot{\phi}}=\pi_{\phi}=\phi'.\label{TMOM3}
\end{equation}
The equation (\ref{TMOM1}) and (\ref{TMOM3}) are  the two primary
constraint of the theory. The effective Hamiltonian (according to
the terminology of Dirac) therefore is given by
\begin{equation}
H_{EF}=\int dx[H_{c}+u\pi_{0}+v(\pi_{\phi}-\phi^\prime)]
\end{equation}
where the canonical Hamiltonian  $H_{CH}$ reads
\begin{equation}
H_{CH} =\int dx[\frac{1}{2}\pi_{1}^{2}+\pi_{1} A'_{0}
+\phi'^2-2e(A_{0}-A_{1})\phi'+2e^{2} A_{1}^{2}]
\end{equation}
The Lagrange multiplier $u$ and $v$ are  are found out in due
course. The Gauss law constraints of this theory is found out to
be
\begin{equation}
G=\pi'_{1}+2e\phi^{\prime}\approx 0. \label{GAUSS}
\end{equation}
from the preservation of the constraint (\ref{TMOM1}).  The
preservation of constraint $(\pi_{\phi}-\phi^{\prime})\approx 0$
with respect to the Hamiltonian gives a new constraint
\begin{equation}
(A_{1}+A_{0})^{\prime}\approx 0. \label{TERCON}
\end{equation}
The Lagrangian multiplier u and v are  given by
\begin{equation}
u=-\pi_{1}+A'_{0},
\end{equation}
\begin{equation}
v=\phi-e(A_{0}-A_{1}).
\end{equation}
Imposing the constraints (\ref{TMOM1}),  (\ref{TMOM3}) and
(\ref{TERCON}) we obtain the reduced Hamiltonian:
\begin{equation}
H_R= \int dx
[\frac{1}{2}\pi_1^{2}+\pi_{1}A'_{0}+\frac{1}{4e^2}\pi'^{2}_1+
4e^{2}A_1^{2}]
\end{equation}
The phase space of the system are  endowed with following four
constraints.
\begin{equation}
\Omega_{1} = \pi_{0} \approx 0,
\end{equation}
\begin{equation}
\Omega_{2}=\pi_{\phi}-\phi^{\prime}\approx 0,
\end{equation}
\begin{equation}
\Omega_{3}=\pi_{1}+2e\phi\approx 0,
\end{equation}
\begin{equation}
\Omega_{4} =A_{1}+A_{0}\approx 0.
\end{equation}
It is constrained theory so Poisson bracket become insufficient to
get the correct equations of motion.  The Dirac brackets
\cite{DIR} give the correct equations of motion which is defined
by.
\begin{equation}
[A(x),B(y)]^{*}
=[A(x),B(y)]-\int[A(x),\Omega_{i}(\eta)]C_{ij}^{-1}[\Omega_{i}(\eta),B(x)]d\eta
dz \label{DEFD}
\end{equation}
where $ C_{ij}^{-1} $ is obtained using the following relation
\begin{equation}
\int C_{ij}^{-1} (x,z)[\omega_{i}(z),\omega_{j}(y)] dz =1
\end{equation}
The Dirac brackets between the fields describing the reduced
Hamiltonian are computed to be
\begin{equation}
[A_1,\pi_{1}]^* = \delta(x-y), \label{DB1}
\end{equation}
\begin{equation}
[A_{1},A_{1}]^*=-\frac{1}{2e^2}\delta^{\prime}(x-y).\label{DB2}
\end{equation}
 The following two first order equations of motion are followed
 from the reduced Hamiltonian with the use of the Dirac brackets
 (\ref{DB1}),and (\ref{DB2}):
\begin{equation}
\dot{\pi_{1}} = \pi'_1 - 4e^2A_{1},
\end{equation}
\begin{equation}
\dot{A_{1}} = \pi_{1} - A'_{1}.
\end{equation}
These two equations lead to two second order differential
equations
\begin{equation}
[\Box + 4e^2]A_1 = 0. \label{SPEC1}
\end{equation}
\begin{equation}
[\Box + 4e^2]\pi_1 = 0.\label{SPEC2}
\end{equation}
Equation (\ref{SPEC1}) represents a massive boson with square of
the mass $4e^2$ and equation (\ref{SPEC1}) stands for the momentum
of the massive field $A_1 $. Let us now look back carefully to the
spectrum obtained in Sec. II through the Lagrangian formulation.
Note that the field $F$ can expressed in terms of the field $A_0$
and $A_1$ by the expression
\begin{equation}
 \pi_1 = -\epsilon_{\mu\nu}\partial^\nu A_\mu = \Box F\approx F
\end{equation}
So the field $F$ corresponds to the field $\pi$. In Sec. II we
 have imposed a condition $A_0 + A_1 = 0$ in an ad hoc manner. But
 the Hamiltonian formulation shows that the condition $A_0 + A_1 = 0$
 is lying hidden within the system which manifests itself
as a constraint in the phase space of the theory.
\section{Determination of theoretical spectrum of the gauge invariant theory in
the Lagrangian
formulation} In this section we extend the phase space of the
theory introducing some new fields following Stuckelberg
formalism. The theory is made gauge invariant putting  the
Wess-Zumino term ${\cal L}_{WZ}$ within the Lagrangian ${\cal
L}_{CH}$
\begin{equation}
{\cal L}_{GS}={\cal L}_{CH}+ {\cal L}_{WZ} \label{GSL},
\end{equation}
where $L_{WZ}$ stands for Wess-Zumino term \cite{WESS}:
\begin{equation}
{\cal L}_{WZ}=-\dot{\zeta}\zeta'-\zeta'^{2}+2e(A_{0}+A_{1})\zeta'.
\end{equation}
It is straightforward to examine that the Lagrangian (\ref{GSL})
is invariant under the transformation $A_\mu\rightarrow A_\mu
+\frac{1}{e}\partial_\mu \lambda$, $\phi\rightarrow\phi+\lambda$
and $\zeta\rightarrow\zeta-\lambda$. In order to determine the
physical specter we need to introduce the gauge fixing condition
which is offered here by the following Lagrangian. So the
Lagrangian with which we will proceed here is
\begin{equation}
{\cal L}_{T}={\cal L}_{CH}+{\cal L}_{WZ}+{\cal L}_{GF},
\label{GFL}
\end{equation}
where
\begin{equation}
{\cal L}_{GF}=B\partial_{\mu}A^{\mu}+\frac{1}{2}\alpha{B^{2}}
\end{equation}
The Euler-Lagrange equations of motion that follows from the
Lagrangian (\ref{GFL}) are the following.
\begin{equation}
-\ddot{A}_1+ \dot{A}'_0-2e\phi'-4e^2A_1 -2\zeta'=0, \label{EQM1}
\end{equation}
\begin{equation}
A''_0+ \dot{A}_1'+2e\phi' + 2\zeta'=0,\label{EQM2}
\end{equation}
\begin{equation}
\partial_-\phi'+e(A_{0}-A_{1})=0,\label{EQM3}
\end{equation}
\begin{equation}
\partial_+\zeta'-e(A_{0}-A_{1})=0,\label{EQM4}
\end{equation}
\begin{equation}
\partial_{\mu}A^{\mu}+\alpha{B}=0.\label{EQM5}
\end{equation}
The ansatz for the fields which are found to be appropriate to
solve the above set of coupled differential equations are
\begin{equation}
A_{\mu}=\frac{1}{4e^{2}}\tilde{\partial}F+\partial_{\mu}B+\partial_{\mu}\chi,
\label{ANT1}
\end{equation}
\begin{equation}
\zeta=-\frac{1}{4e}F-B-\chi, \label{ANT2}
\end{equation}
\begin{equation}
\phi=\frac{1}{4e}F-B-\chi. \label{ANT3}
\end{equation}
Using the ansatz (\ref{ANT1}), \ref{ANT2}) and \ref{ANT3}) in the
equations of motion (\ref{EQM1}), \ref{EQM2}), \ref{EQM3}),
\ref{EQM4}) and \ref{EQM5}) we obtain the following three second
order differential equations:
\begin{equation}
(\Box+4e^{2})\Box{F}=0, \label{1EQ}
\end{equation}
\begin{equation}
\Box{B}=0, \label{2EQ}
\end{equation}
\begin{equation}
\Box{\chi}+\alpha{B}=0,\label{3EQ}
\end{equation}
where
\begin{equation}
\pi_{1}=\dot{A_{1}}-A'_{0}=\frac{\Box{F}}{4e^{2}}
\end{equation}
The field $F\approx \Box F$ is representing the massive field with
mass $2e$. The corresponding equation we have obtained in both the
Lagrangian and Hamiltonian formulation of the theory with its
gauge non-invariant version in equation (\ref{OS}) and
(\ref{SPEC1}) or (\ref{SPEC2}) respectively. The equation
(\ref{2EQ}) appears because in the gauge fixed Lagrangian we have
used an auxiliary field $B$ and  the field $\chi$ represents the
zero mass dipole field playing the role of gauge degrees of
freedom that can be eliminated by operator gauge transformation.
So the spectrum agrees with spectrum obtained in Sec. III.
\section{To make an equivalence between the gauge invariant and
 gauge non-invariant version}
To make an equivalence between the gauge invariant and the gauge
non-invariant version of this model we proceed with the gauge
symmetric Lagrangian. So we add up the Wess-Zumino term with the
usual Lagrangian.
\begin{equation}
L_{GS}=\int dx [{\cal L}_{CH}+{\cal L}_{WZ}]
\end{equation}
\begin{eqnarray}
L_{GS}&=&\int dx
[\dot{\phi}\phi'-\phi'^{2}+2e(A_{0}-A_{_1})-2e^{2}A_{1}^{2}
-\dot{\zeta}\zeta'\nonumber
\\ &-&\zeta'^{2}
+2e(A_{0}+A_{1})\zeta'+\frac{1}{2}(\dot{A}_1-A'_0)^2]\label{LGS}
\end{eqnarray}
Equations of motion are
\begin{equation}
\frac{\partial L_{GS}}{\partial\dot{\phi}}=\pi_{\phi}=\phi',
\label{MOMS1}
\end{equation}
\begin{equation}
\frac{\partial L_{GS}}{\partial\dot{\zeta}}=\pi_{\zeta}=-\zeta',
\label{MOMS2}
\end{equation}
\begin{equation}
\frac{\partial L_{GS}}{\partial\dot{A}_0}=\pi_0\approx 0,
\label{MOMS3}
\end{equation}
\begin{equation}
\frac{\partial
L_{GS}}{\partial\dot{A}_1}=\pi_1=(\dot{A}_1-A'_0).\label{MOMS4}
\end{equation}
By the use of equations (\ref{MOMS1}), (\ref{MOMS2}),
(\ref{MOMS3}) and (\ref{MOMS4}) thge canonical Hamiltonian that
follows from the Lagrangian (\ref{LGS}) is
\begin{equation}
H_{CGS}=\int dx[
\pi_{1}\dot{A_{1}}+\pi_{0}\dot{A_{0}}+\pi_{\phi}\dot{\phi}+\pi_{\zeta}\dot{\zeta}]-L_{GS}
\end{equation}
Therefore, the effective Hamiltonian for this situation  is
\begin{eqnarray}
H_{GSE}&=&\int dx[\frac{1}{2}\pi_{1}^{2}+\pi_{1}A'_{0}+\phi'-2e(A_{0}-A_{1})\phi'
+2e^{2}A_{1}^{2}+\zeta'^{2} \nonumber \\
&-&2e(A_{0}+A_{1})\zeta'
+u\pi_{0}+v(\pi_{\phi}-\phi')+w(\pi_{\zeta}+\zeta')]
\end{eqnarray}
The Gauss law constraint that comes out from the preservation
condition of constraint (\ref{MOMS3}) is
\begin{equation}
G=[\pi_{0},H]=\pi'_{1}+2e(\phi'+\zeta').
\end{equation}
The velocities $u$ and $v$ are found out as
\begin{equation}
v=\phi'+e(A_{0}-A_{1}),
\end{equation}
\begin{equation}
w=-\zeta'+e(A_{0}+A_{1}).
\end{equation}
Using the velocities and the successive use of the condition of
preservation of the constraints the following set of second class
constrain are found to be embedded within the phase space of the
theory.
\begin{equation}
{\cal C }_{1}=\pi_{0}\approx 0
\end{equation}
\begin{equation}
{\cal C }_{2}=\pi_{\phi}-\phi'\approx 0
\end{equation}
\begin{equation}
{\cal C }_{3}=\pi_{\zeta}+\zeta'\approx 0
\end{equation}
\begin{equation}
{\cal C }_{4}=\pi'_{1}+2e(\phi'+\zeta')\approx 0
\end{equation}
We are in a position to chose gauge fixing conditions those which
are very crucial in this situation. The inappropriate use of gauge
fixing  leads to different effective theory which may mislead to
reach to the goal. The appropriate gauge fixing which meets our
need are the following.
\begin{equation}
{\cal C }_{5}=\zeta'=0,
\end{equation}
\begin{equation}
{\cal C }_{6}=\pi_{\zeta}=e(A_{0}+A_{1}).
\end{equation}
These inputs  therefore enables us to write down the generating
functional.
\begin{eqnarray}
Z&=& \int[det[{\cal C }_{l},~{\cal C
}_{m}]]^{\frac{1}{2}}dA_{1}d\pi_{1} d\phi d\pi_{\phi}
dA_{0}d\pi_{0} d\zeta d\pi_{\zeta}e^{i\int
d^2x(\pi_{1}\dot{A_{1}}+\pi_{\phi}\dot{\phi}+\pi_{\zeta}\dot{\zeta}
+\pi_{0}\dot{A_{0}}-H_{C})}\nonumber\\
&&\times \delta({\cal C }_{1})\delta({\cal C }_{2})\delta({\cal C
}_{3})\delta({\cal C }_{4})\delta({\cal C }_{5})\delta({\cal C
}_{6}). \label{GENF}
\end{eqnarray}
Here $l$ and $m$ runs from $1$ to $4$. Integrating out of the
fields $\zeta$ and $\pi_{\zeta}$ we find that equation
(\ref{GENF}) reduces to
\begin{eqnarray}
Z&=& \int dA_{1}d\pi_{1} d\phi d\pi_{\phi}dA_{0}d\pi_{0}e^{i\int
d^2x(\pi_{1}\dot{A_{1}}+\pi_{\phi}\dot{\phi}
+\pi_{0}\dot{A_{0}}-{\cal H}_{GSF})}\nonumber\\
&&\times \delta({\cal C }_{1})\delta({\cal C }_{2})\delta({\cal C
}_{3})\delta({\cal C }_{4}).
\end{eqnarray}
where
\begin{equation}{\cal
H}_{GSF}=\frac{1}{2}\pi_{1}^{2}+\pi_{1}A'_{0}+\phi'-2e(A_{0}-A_{1})\phi'+2e^{2}A_{1}^{2}
\end{equation}
Again integrating out of the momenta $\pi_0$, $\pi_1$ and
$\pi_\phi$ leads us to
\begin{equation}
Z=\int d\phi dA_{1}dA_{0} e^{i\int d^2x {\cal L}_{GSF}}
\end{equation}
where
\begin{equation}
{\cal
L}_{GSF}=\dot{\phi}\phi'-\phi'^{2}+2e\phi'(A_{0}-A_{1})-2e^2A^{2}_{1}
+\frac{1}{2}(\dot{A_{1}}-A'_{0})^{2}\end{equation} Note that the
system now contains the usual four constraint ${\cal C
}_{1}$,${\cal C }_{2}$,${\cal C }_{3}$,${\cal C }_{4}$ and
$L_{GSF}$ is the identical to usual Lagrangian $L_{CH}$ having the
same Hamiltonian $H_{GSR}= H_{R}$. So the gauge invariant
Lagrangian maps on to gauge non-invariant Lagrangian of the usual
phase space. It also ensures that the physical contents in both
the version are identical.

\section{Study of BRST symmetry of the model }
Let us now turn towards the study of BRST symmetry of the
effective action. It is an important symmetry which ensures the
unitarity and renormalization  of a theory \cite{BRS1, BRS2,
BRS3}. In the articles \cite{BAT1, BAT2, BAT3, BAT4, FIK}
different method of construction of BRST invariant effective
action of a given theory have been discussed and the applications
of these formalism have been pursued in the articles \cite{KIM1,
KIM2, KIM3, MIAO, AR1, AR2, AR3}. Since BRST  is a powerful tools
to ensure the the unitarity and renormalization  of a theory the
study of BRST symmetry of any  field theoretical model would be of
interest. In this section we are intended to study the BRST and
anti-BRST properties of this model both in off-shell and on-shell
environment. With out going through the formal construction of
BRST invariant effective action we discuss here the symmetry
properties of this model with different gauge fixing term. To this
end we consider gauge invariant version of the action with the
Lorentz gauge fixing term
\begin{equation}
S_{BRST}=\int
d^2x[L+L_{WZ}+\partial_{\mu}A^{\mu}B+\frac{1}{2}\alpha B^{2}
+\partial_{\mu}\bar{C}\partial^{\mu}C]
\end{equation}
It is straightforward to see that the Lagrangian is off-shell
invariant under the  BRST transformation
\begin{equation}
\delta_{B}A_{\mu}=-\frac{1}{e}\lambda\partial_{\mu}C,~~\delta_{B}\phi=\lambda
C , ~~\delta_{B}\zeta=-\lambda C
\end{equation}
\begin{equation}
\delta_{B}\bar{C}=\lambda B,~~\delta_{B}C=0,~~\delta_{B}B=0.
\end{equation}
The above Lagrangian is found to be off-shell invariant under the
anti-BRST transformation
\begin{equation}
\delta_{ab}A_{\mu}=-\frac{1}{e}\lambda\partial_{\mu}\bar{C},
~~\delta_{ab}\phi=\lambda\bar{C},~~\delta_{ab}\zeta=-\lambda\bar{C}
\end{equation}
\begin{equation}
\delta_{ab}C=\lambda B,~~\delta_{ab}\bar{C}=0,~~\delta_{ab}B=0.
\end{equation}
The gauge fixing condition can be chosen in different ways keeping
the physical contents intact. We choose another important gauge
fixing term which is known as 't Hooft-Veltman gauge. The
effective action with this gauge reads
\begin{equation}
\bar{S}_{BRST}=\int d^2x [{\cal L} +{\cal
L}_{WZ}+B(\partial_{\mu}A^{\mu}+ eA_\mu A^\mu) +\frac{1}{2}\alpha
B^{2} +\bar{C}(\Box+ e A_\mu\partial^{\mu})C].
\end{equation}
This new effective action is found to be  off-shell invariant with
the following BRST transformation
\begin{equation}
\delta_{B}A_{\mu}=-\frac{1}{e}\lambda\partial_{\mu}C,~~\delta_{B}\phi=\lambda
C , ~~\delta_{B}\zeta=-\lambda C,
\end{equation}
and the above effective action is off-shell invariant under the
following anti-BRST transformation
\begin{equation}
\delta_{ab}A_{\mu}=-\frac{1}{e}\lambda\partial_{\mu}\bar{C},
~~\delta_{ab}\phi=\lambda\bar{C},~~\delta_{ab}\zeta=-\lambda\bar{C}
\end{equation}
\begin{equation}
\delta_{ab}C=\lambda B,~~\delta_{ab}\bar{C}=0,~~\delta_{ab}B=0.
\end{equation}
Let us see whether the gauge fixing condition can be chosen in
such away when the effective action shows  on-shell invariance
under the BRST and anti-BRST transformation. To this end we
consider the following effective action with a different gauge
fixing term.
\begin{equation}
\tilde{S}_{BRST}=\int d^2x[L+L_{wz}+\frac{1}{2\alpha}(\partial.A
+\alpha eB)^{2}
+\partial_{\mu}\bar{C}\partial^{\mu}C+e^{2}\bar{C}C].
\end{equation}
Under the BRST transformation
\begin{equation}
\delta_{b}A_{\mu}=-\frac{1}{e}\lambda\partial_{\mu}C,~~
\delta_{b}\zeta=-\lambda C,~~ \delta_{b}\phi=\lambda C
\end{equation}
\begin{equation}
\delta_{b}C=0,~~ \delta_{b}\bar{C}=\partial.A +\alpha eB,~~
\delta_{b}(\partial.A+\alpha eB)=(\Box + e^{2})C,
\end{equation}
the effective theory is on-shell BRST invariant with the on-shell
 condition
\begin{equation}
(\Box+e^{2})C=0
\end{equation}
The effective theory is also found to be on-shell invariant under
the anti-BRST Transformation
\begin{equation}
\delta_{ab}A_{\mu}=-\frac{1}{e}\lambda\partial_{\mu}\bar{C},~~
\delta_{ab}\zeta=-\lambda \bar{C},~~\delta_{ab}\phi=\lambda
\bar{C}
\end{equation}
\begin{equation}
\delta_{ab}\bar{C}=0,~~ \delta_{ab}C=\partial.A +\alpha eB,~~
\delta_{ab}(\partial.A+\alpha eB)=(\Box + e^{2})\bar{C}
\end{equation}
with the  on-shell condition
\begin{equation}
(\Box+e^{2})\bar{C}=0
\end{equation}
This is in short BRST and anti-BSRT symmetric property of this
theory in the off-shell and on-shell domain.

\section{Summary and discussion}
We have considered a gauged Lagrangian with a Siegel type chiral
boson with the different masslike term for gauge fields. The
masslike term which was chosen in \cite{BEL} led to a gauged
theoryof Florenini-Jackiw type chiral boson which can be derived
from Chiral Schwinger model with the Jackiw-Rajaraman type of
electromagnetic anomaly \cite{KH}. An alternative masslike term is
chosen here in order to derive the gauged model Florenini-Jackiw
type chiral boson which gets generated from the chiral Schwinger
model with Faddeevian type of anomaly \cite{ARC, AR1}. In the
article \cite{PM} the author showed that the Chiral Schwinger
model remains physically sensible in all respect with an
independent type of masslike term where the nature of the anomaly
belonged to the Faddeevian class.

This physical spectrum of the gauge non-invariant version of this
model is found out both in the Lagrangian and Hamiltonian
formulation. We should mention here that the determination of the
spectrum in the Hamiltonian formulation was done in \cite{PM}, but
in the Lagrangian formulation, it was lacking. Here we have done
it in a much transparent manner that enables us to correlate the
spectrum of the theory in both the formulation. It was found that
a condition $A_0 + A_1 =0$ needed to be put in an ad-hoc manner to
obtain the correct spectrum in the Lagrangian formulation. Though
it seems to be unnatural a critical review shows that the
condition $A_0 + A_1 =0$ is basically a constraint of the theory
which shows its mysterious appearance in the Hamiltonian
formulation. In the article \cite{ARS} the similar type analysis
was pursued for the bosonised version of chiral Schwinger model
with the usual Jackiw-Rajaraman type of anomaly. In this work, we
have extended it for the Chiral Schwinger model with Faddeevian
anomaly proposed by Mitre in \cite{PM, PMS}. These model are very
much different so far constraint structure and the physical
spectrum is concerned. So this work though looks similar to
\cite{ARS} as the computation is concerned it will shed light in
the lower dimensional constrained field theoretical regime. For
instance, the solvability of Lagrangian formulation here needs an
extra condition which has appeared as a constraint in the
Hamiltonian formulation. Without that condition, this model would
not be solvable in the Lagrangian formulation.

The model is made gauge-invariant with the incorporation of
Wess-Zumino field. The phase space determination of the model is
then carried out in the Lagrangian formulation using Lorentz gauge
condition since with this gauge the model does not lose its
exactly solvable nature. The theoretical spectrum is found to be
in exact agreement with its gauge non-invariant counterpart. The
auxiliary fields found to remains allocated in the un-physical
sector of the theory. An attempt is made to make an equivalence to
the theory of the gauge-invariant and gauge non-invariant version
using the ingenious formalism developed in \cite{FALCK}. It is
found the role of gauge fixing is very crucial here.

BRST symmetry related to this model is also studied with different
gauge fixing terms. A gauge fixing term has been used where the
model shows on-shell gauge invariance.

\section{Acknowledgements}
 AR likes to thanks  the Director of Saha Institute of Nuclear
Physics, Kolkata, for providing  the computer and library
facilities of the Institute.

\end{document}